%% file: 00-root.tex
\documentclass[conference,compsoc]{IEEEtran}

\input{01-packages}

\input{03-title}

\begin{document}
\maketitle
\thispagestyle{plain}
\pagestyle{plain}

\input{05-abstract}
\input{12-diagram}
\input{10-intro}

\input{20-requisite-math}

\input{30-results}

\input{31-image}

\input{40-conclusions}

\bibliographystyle{IEEEtran}
\bibliography{references}

\end{document}

%% file: 01-packages.tex
\usepackage[dvipsnames]{xcolor}

\IEEEoverridecommandlockouts 

\usepackage{amsmath,amssymb,amsfonts}
\usepackage{mathtools}
\usepackage{bbold}
\usepackage{algorithmic}
\usepackage{graphicx}
\usepackage{tikz}
\usetikzlibrary{arrows.meta, positioning, shapes.geometric}
\usepackage{textcomp}
\usepackage{csquotes}
\usepackage{soul}
\usepackage{hyperref}
\usepackage[hang,flushmargin]{footmisc} 
\hypersetup{
  colorlinks   = true, 
  urlcolor     = blue, 
  linkcolor    = blue, 
  citecolor   = red 
}
\usepackage[]{cite}
\usepackage{verbatim}
\def\BibTeX{{\rm B\kern-.05em{\sc i\kern-.025em b}\kern-.08em 
T\kern-.1667em\lower.7ex\hbox{E}\kern-.125emX}}

\newcommand{\C}{\mathbb{C}}

\newcommand{\R}{\mathbb{R}}

%% file: 03-title.tex
\title{\Large \bf
Destabilizing a Social Network Model via Intrinsic Feedback Vulnerabilities
}

\author{Lane H. Rogers$^{1}$, Emma J. Reid$^{2}$, and Robert A. Bridges$^{3}$%
\thanks{Notice: This manuscript has been authored by UT-Battelle, LLC, under contract DE-AC05-00OR22725 with the US Department of Energy (DOE). The US government retains and the publisher, by accepting the article for publication, acknowledges that the US government retains a nonexclusive, paid-up, irrevocable, worldwide license to publish or reproduce the published form of this manuscript, or allow others to do so, for US government purposes. DOE will provide public access to these results of federally sponsored research in accordance with the DOE Public Access Plan (http://energy.gov/downloads/doe-public-access-plan).}%
\thanks{$^{1}$Lane H. Rogers is with University of Tennessee, Knoxville, Department of Mathematics, 1403 Circle Dr, Knoxville, TN 37916, USA
        {\tt\small lrogers@tennessee.edu}}%
\thanks{$^{2}$Emma J. Reid is with Oak Ridge National Laboratory, 1 Bethel Valley Road Oak Ridge, TN 37830, USA
        {\tt\small reidej@ornl.gov}}%
\thanks{$^{3}$Robert A. Bridges is with AI Sweden, Lindholmspiren 11, 417 56 Göteborg, Sweden
        {\tt\small robert.bridges@ai.se}}%
}

%% file: 05-abstract.tex
\begin{abstract}
Social influence plays a significant role in shaping individual sentiments and actions, particularly in a world of ubiquitous digital interconnection. 
The rapid development of generative artificial intelligence (AI) has given rise to well-founded concerns regarding the potential implementation of radicalization techniques in social media. 
Motivated by these developments, we present a case study investigating the effects of small but intentional perturbations 
on a simple social network. 
We employ Taylor’s classic model of social influence and tools from robust control theory (most notably the Dynamical Structure Function (DSF)), to identify perturbations that qualitatively alter the system's behavior while remaining as unobtrusive as possible. 
We examine two such scenarios: perturbations to an existing link and perturbations that introduce a new link to the network.
In each case, we identify destabilizing perturbations of minimal norm and simulate their effects. 
Remarkably, we find that 
small but targeted alterations to network structure may lead to the radicalization of all agents, exhibiting the potential for large-scale shifts in collective behavior to be triggered by comparatively minuscule adjustments in social influence. 
Given that this method of identifying perturbations that are innocuous yet destabilizing applies to \textit{any} suitable dynamical system, our findings emphasize a need for similar analyses to be carried out on real systems (e.g., real social networks), to identify the places where such dynamics may already exist. 

\end{abstract}

%% file: 12-diagram.tex
\begin{figure}[htbp] 
\centering 
\caption{Flowchart of perturbation framework: \textcolor{NavyBlue}{(1)} a given ODE system model is restricted to only the exposed variables (states vulnerable to adversarial observation or perturbation); \textcolor{NavyBlue}{(2)} taking the Laplace transform reveals the transfer function $G$, which is bounded if 
the system is asymptotically stable; \textcolor{NavyBlue}{(3)} the DSF enables the computation of a minimally-normed perturbation, $\Delta(s)$, guaranteed to destabilize the system; \textcolor{NavyBlue}{(4)} inverting the Laplace transform allows for a return to the time domain to verify instability.}


\vspace{.25cm}

\label{fig:flow-chart}
\begin{tikzpicture}[node distance=1.2cm, auto, every text node part/.style={align=center}, scale=0.68, transform shape] 

  \node (Y1) [draw, rectangle,  minimum width=3.5cm, minimum height=1cm] {$Y(s) = G(s) U(s)$};

  \node (model) [draw, rectangle, below=of Y1, minimum width=3.5cm, minimum height=1cm] { 
  $\dot{y}(t) = Ay(t) + Bu(t)$}; 
\node[circle, fill = NavyBlue, text = white,  minimum size=0.35cm] at ([xshift=-1.75cm]model.north) {1};

\node[circle, fill = NavyBlue, text = white,  minimum size=0.35cm] at ([xshift=-1.75cm]Y1.north) {2};

  \node [above=0.15cm of Y1]
  {\textit{\textbf{Original System}}};

    \node(plot1)[below=of model, yshift=.5cm]{
        \begin{tikzpicture}[scale=.9]
        \draw[->] (0,0) -- (3,0) node[right] {$t$};
        \draw[->] (0,0) -- (0,1.7) node[above] {$y_{i}$};
        \draw[Green, thick, smooth, samples=100, domain=0:3] plot (\x, {.7 - .3 * 4*sin(deg(5*1.3*\x)) * exp(-3.5*1.3*\x)});
        \draw[violet, thick, smooth, samples=100, domain=0:3] plot (\x, {.5 - .2 * 2*cos(deg(5*1.3*\x)) * exp(-2.5*1.3*\x)});
        \end{tikzpicture}
    };

  \draw[->, >=Stealth] (model) to node[right, midway, font=\small] {Laplace \\ transform
  } (Y1); 

  \draw[->, >=Stealth] (model) to node[left, midway, font=\small] {} (plot1); 

    \node (Y2) [draw, rectangle, right=of Y1, xshift=2.5cm, minimum width=4.25cm, 
    minimum height=1cm] 
    {$\scriptstyle Y(s) = [I - H(s)\textcolor{BrickRed}{\Delta(s)}]^{-1} G(s)U(s)$};
    \node[circle, fill = NavyBlue, text = white,  minimum size=0.35cm] at ([xshift=-2.1cm]Y2.north) {3};

    \node (model-pullback) [draw, rectangle, below=of Y2, minimum width=4.25cm, minimum height=1cm] {$\scriptstyle y'(t) = (A\textcolor{BrickRed}{+\Delta(t)})y(t)+Bu(t)$};
    \node[circle, fill = NavyBlue, text = white,  minimum size=0.35cm] at ([xshift=-2.1cm]model-pullback.north) {4};
    
    \node(plot2)[below=of model-pullback, yshift=.5cm]{
        \begin{tikzpicture}[scale=.9]
        \draw[->] (0,0) -- (3,0) node[right] {$t$};
        \draw[->] (0,0) -- (0,1.7) node[above] {$y_{i}$};
        \draw[Green, thick, smooth, samples=100, domain=0:3] plot (\x, {.2 - .3 * 4*sin(deg(5*\x))* exp(-3*\x) + exp(.5*\x-.9)*.8});
        \draw[violet, thick, smooth, samples=100, domain=0:3] plot (\x, {0 - .2 * 2*cos(deg(5*\x)) * exp(-2.5*\x)+ exp(.4*\x-.9)});
        \end{tikzpicture}
        };
    \node [above=0.15cm of Y2] {\textit{\textbf{Perturbed System}}};

  \draw[->, >=Stealth] (Y2) to 
    node[right, midway, font=\small] {inverse Laplace \\ transform } (model-pullback); 
  \draw[->, >=Stealth] (model-pullback) to node[right, midway, font=\small] {} (plot2); 

 \draw[->, >=Stealth, BrickRed, line width=0.8pt] (Y1) to node(red)[above, midway, font =\scriptsize]{introduce $\textcolor{BrickRed}{\Delta}(s)$} (Y2);
\end{tikzpicture}
\end{figure}

%% file: 10-intro.tex
\section{Introduction\texorpdfstring{\protect\footnote[4]{These results may be reproduced freely at \url{https://github.com/lane-h-rogers/Social_Network_Destabilization}.}}{}} \label{sec:intro}
Social influence refers to the ways in which the sentiments and actions of an individual are affected by both social interaction and content from information feeds \cite{cialdini2004social}.
Recent technological advances have expanded individual 
spheres of social influence far beyond mere in-person interaction. 
As such methods of social influence evolve, it is imperative to evaluate the impact they have on individual and collective sentiments.
While AI demonstrates promise in bolstering the integrity of news delivered via social media, it has also become an increasingly 
central tool for delivering highly tailored or misleading content to a particular individual's insular feed of information \cite{kreps2022all, horne2019rating}.
Individual cases of socially engineered radicalization resulting from algorithmic promotion of incendiary content have even broached the United States Supreme Court \cite{taamneh2023case, gonzalez2023case}. 
%
%

In this paper, we present a case study of a simple social influence model to determine whether or not small perturbations are indeed capable of producing significant change in the long-term disposition of several network agents. 
To this end, we employ Taylor's model of social influence \cite{taylor1968towards}, in which the sentiment of multiple agents evolves according to both the influence the agents exert over one another, as well as the presence of external sources such as mass media. 
A simulation of the long-term behavior of the social network reveals convergence to a stable equilibrium of dissenting sentiments.
Robust control theory \cite{dullerud2013course} provides the mathematical framework to assess the robustness of the system to perturbation.
This provides a rigorous method for quantifying the ``magnitude'' of vulnerability 
and identifies a stability threshold for such perturbations. 
In particular, we consider two classes of perturbations. 
``Existing-link'' perturbations restricts alterations to a single existing link in the social influence graph, meaning that no influence may be introduced where none already exists.
In contrast, ``created-link'' perturbations expand the set of possible alterations to include the introduction of a single new link of influence to the social network where none previously existed.
In both cases, we use the Dynamical Structure Function (DSF) \cite{chetty2014vulnerability, debuse2024study, GoncalvesDSF} 
to identify a destabilizing perturbation of minimal norm and simulate the 
perturbed system.
These techniques generalize to \textit{any} dynamical system model satisfying a general set of hypotheses. 

In each perturbed case, the resulting change in long-term system behavior is not slight: 
the sentiments of all agents in the perturbed networks grow without bound. 
In the case of this particular social influence structure, our findings indicate that a small, persistent, and targeted perturbation can cause the radicalization of all agents.
As such, it serves as motivation to identify such vulnerabilities and subsequently reinforce social networks against them.

%% file: 20-requisite-math.tex
\section{Requisite Background}
In this section, we introduce Taylor's model of social network dynamics and provide a brief overview of the DSF and its role in the vulnerability analysis of dynamical systems. We refer readers to prior works for the many details that elude the current scope \cite{dullerud2013course, rai2012vulnerable, grimsman2016case, GoncalvesDSF, gonccalves2008necessary}.
\subsection{Taylor's Social Network Model}\label{sec:taylor-model}
\begin{figure}[t]
    \centering
    \includegraphics[width=0.375\textwidth]{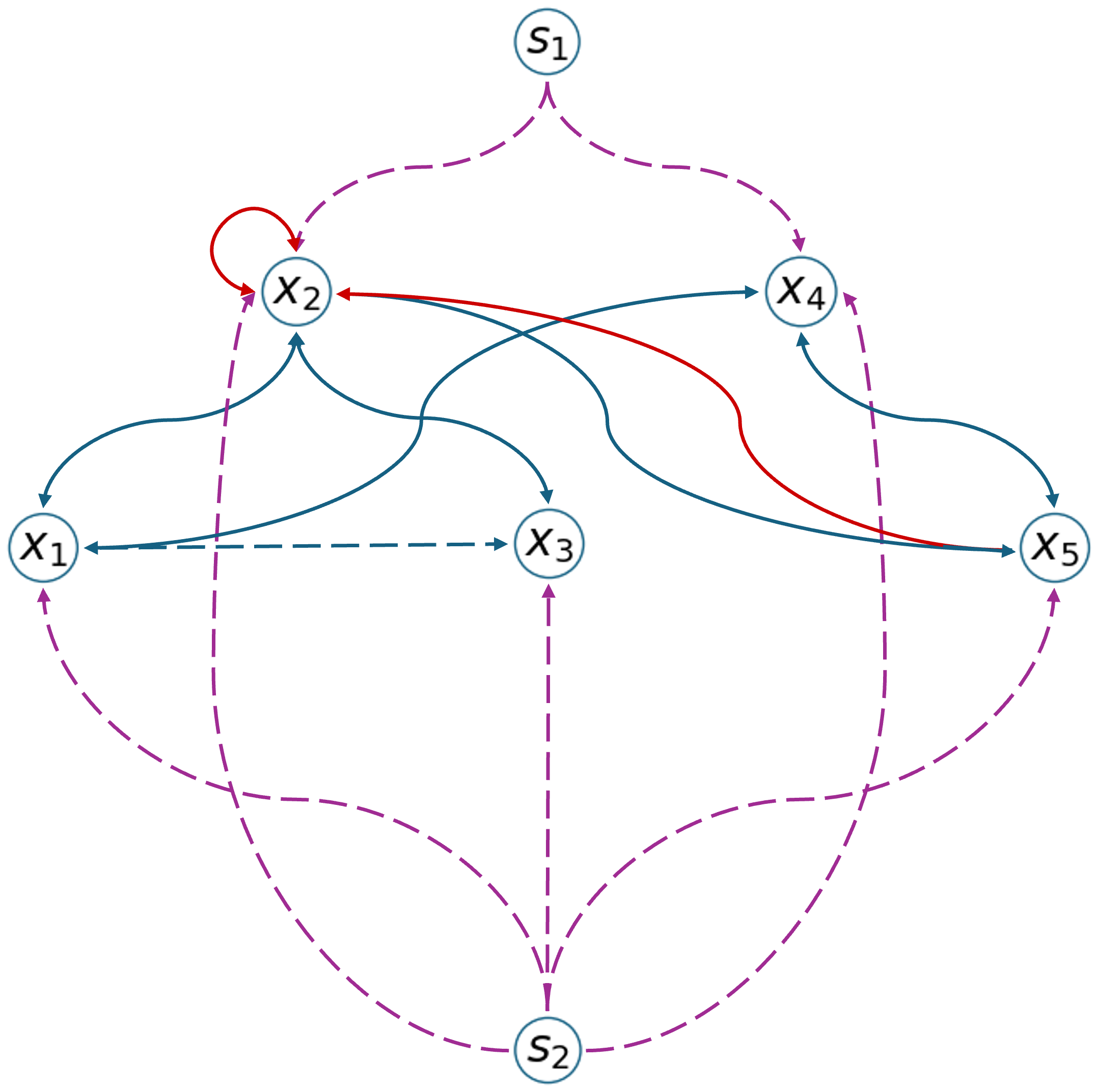}
    \caption{Graph-theoretic representation of a simple social network with five distinct agents and two distinct static sources. ``One-way'' influence is denoted with a dashed edge, while reciprocal influence (which may still be asymmetric in magnitude) is denoted by a solid edge. Source-influence is colored purple, while agent-influence is colored blue. In red, we indicate the most vulnerable links of agent-influence (both pre-existing and created) as discussed in Sections (\ref{sec:existinglink}) and (\ref{sec:createdlink}). See Figure (\ref{fig:comparison}) for a comparison of the simulated system effects with and without the perturbations.}
    \label{fig:network}
\end{figure}
Building on the continuous-time model of social influence by Abelson \cite{abelson1967mathematical}, Taylor's model contributes additional realism via the addition of ``stubborn'' agents, or ``sources''. These are agents who, unlike their malleable counterparts, are not prone to external influence of any kind.  Taylor's model was chosen in favor of more primitive models, such as that of Abelson or even French-Degroot, due to both its well-corroborated status as a model of social interaction and its realistic expression of enduring dissent
 \cite{proskurnikov2017tutorial1}.

The model may be summarized mathematically as 
$\dot{x} = -(L + \Gamma)x + \Gamma u.$ 
Here $L$ is the familiar Laplacian matrix of the network represented in Figure (\ref{fig:network}), with weights that signify the influences inherent to this particular network \cite{proskurnikov2017tutorial1}. Here $\Gamma$ is a diagonal matrix satisfying
\[
\gamma_{ii} = \sum_{m=1}^{k} p_{im} \geq 0,
\]
where $p_{im} \geq 0$ are ``persuasibility parameters'' that describe the magnitude of influence that sources $s_{1}, \dots, s_{m}$ have on agent $x_{i}$, respectively. 
This matrix must be nonnegative since no agent is completely free of source influence. Finally, we define
\[
u_{i} = \gamma_{ii}^{-1} \sum_{m=1}^{k} p_{im}s_{m},
\]
where $s_{m}$ are the sentiments of the respective broadcast sources present in the model. The trivial case in which $\gamma_{ii} = 0$ is addressed by also setting $u_{i} = 0$.
The above may be simplified to
\[
\dot{x}(t) = Ax(t) + b
\]
where $x = [x_1, ..., x_n]^T$ is the vector of states $x_i(t)$ quantifying the sentiment of agent $x_{i}$ at time $t$, 
$A$ is $n\times n$ providing the agents' influence on each other, and $b = [b_1, ..., b_n]^T$ the vector of influence from sources.

As mentioned above, more rudimentary social network models \cite{abelson1964mathematical, abelson1967mathematical, proskurnikov2017tutorial1} suffer from a ubiquitous tendency to converge toward a state of total consensus.
Taylor's model allows for a stable equilibrium state of dissenting sentiments via its inclusion of static sources \cite{taylor1968towards}.
The effect of these
is represented above by the inhomogeneous term $b$ that provides the cumulative influence of the static sources on the malleable agents. 
\subsection{Dynamical Structure Function} \label{sec:dsf}
DSFs give rise to a graphical interpretation of the underlying dynamical system, separate from the original graph-theoretic representation), that summarizes the causal relationships between states. We use this representation for vulnerability analysis and consequent destabilization.
Recall that a continuous-time dynamical system of differential equations, \begin{equation}\label{eqn:control-theory-eqn}
\dot{x}(t) = Ax(t) + Bu(t)
\end{equation}
is said to be asymptotically stable (at an equilibrium $x = 0$) 
if $\sigma(A) \subset \C_-$, i.e., all eigenvalues lie in the open left-half complex plane. It is unstable if there exists $\lambda \in \sigma(A) \cap \C_+$ i.e. at least one eigenvalue lies in the open right-half complex plane, and may still be classified as ``marginally'' stable otherwise. 
Robust control theory provides a framework for quantifying resilience of an asymptotic equilibrium state to perturbations and the DSF uses this framework for vulnerability analysis. 
\subsubsection*{Exposed States}
We designate state variables as either ``exposed'' or ``hidden''. For our purposes, exposed state variables are considered susceptible to being both observed and manipulated, whereas hidden state variables are not.
%
Without loss of generality, we assume that the exposed variables constitute the first $p \leq n$ indices of the state vector $x(t) \in \R^{n}$. Denoting the vector of exposed states $y(t) \in \R^{p}$ and the vector of remaining hidden states as $z(t) \in \R^{n-p}$, it is easy to conclude that restricting the analysis to the dynamics of $y(t)$ allows one to observe the system from the point of view of a potential attacker. 
For instance, the control system defined by Equation (\ref{eqn:control-theory-eqn})
may be partitioned into exposed and hidden states as $x(t) = [y(t) \quad z(t)]^T$, which yields
\[
\begin{bmatrix}
    \dot{y}(t) \\
    \dot{z}(t)
\end{bmatrix}
=
\begin{bmatrix}
    A_{11} & A_{12} \\
    A_{21} & A_{22}
\end{bmatrix}
\begin{bmatrix}
    y(t) \\
    z(t)
\end{bmatrix}
+
\begin{bmatrix}
    B_{1} \\
    B_{2}
\end{bmatrix}
u(t).
\]
%
%
\subsubsection*{Computing the DSF} 
To compute the DSF, we begin by taking the Laplace transform of the attack surface model, which yields
\begin{equation}\label{eqn:prepredsf}
\begin{bmatrix}
    sY(s) \\
    sZ(s)
\end{bmatrix}
=
\begin{bmatrix}
    A_{11} & A_{12} \\
    A_{21} & A_{22}
\end{bmatrix}
\begin{bmatrix}
    Y(s) \\
    Z(s)
\end{bmatrix}
+
\begin{bmatrix}
    B_{1} \\
    B_{2}
\end{bmatrix}
U(s),
\end{equation}
an expression in the frequency domain. After some algebra, we obtain 
\begin{align} \begin{split} \label{preDSF}
sY(s) &= \tilde{Q}(s)Y(s) + \tilde{P}(s)U(s), \\
\text{with }\ \ \tilde{Q}(s) &= A_{11} + A_{12}(sI - A_{22})^{-1}A_{21}, \\
\text{and }\ \ \tilde{P}(s) &= B_1 + A_{12}(sI - A_{22})^{-1}B_2.
\end{split}\end{align}
Denote $D(s) = \text{diag}(\tilde{Q})$. Subtracting $D(s)Y(s)$ from each side of Equation (\ref{preDSF}) yields
\begin{align}\begin{split}\label{eqn:dsf}
Y(s) &= Q(s)Y(s) + P(s)U(s), \\
\text{where }\ \ Q(s) &= (sI - D(s))^{-1}(\tilde{Q}(s) - D(s)), \\
\text{and } \ \ P(s)&= (sI - D(s))^{-1}\tilde{P}(s).
\end{split}\end{align}
The ordered pair $(Q(s), P(s))$ is precisely the (unique) DSF, where $Q$ provides the causal influence the exposed states $Y$ have on each other, while $P$ provides the causal influence of the inputs $U$ on the exposed states $Y$. 
%
%
\subsubsection*{Vulnerability Analysis via the DSF}
One upshot of the DSF is it allows simple analysis of a system's stability. In particular, it may be used to determine a perturbation of \textit{minimal magnitude} (using the $\mathcal{H}_\infty$ matrix norm) that would destabilize the system. While there are many potential perturbations that would successfully destabilize the given system, we only seek those that are minimal in $\mathcal{H}_\infty$ norm.

To examine the impact of a destabilizing perturbation, we solve Equation~(\ref{eqn:dsf}) for $Y$, giving $Y=(I-Q)^{-1}PU$. It follows that an expression for the system's transfer function is given by $G = (I-Q)^{-1}P$. Recall that an unbounded transfer function (in the $\mathcal{H}_\infty$ norm) implies that a system is not asymptotically stable \cite{dullerud2013course}. 

Previous efforts in DSF analysis have shown that additive perturbations to $P$ will not destabilize the system, so we need only consider additive perturbations to $Q$ \cite{rai2012vulnerable}.
This corresponds to an additive perturbation to the original ODE system.
We model the perturbed system as $Y = (Q+\Delta)Y +PU$ where addition $\Delta Y$ represents the perturbation of exposed variables.
The perturbed system's transfer function is 
\[
(I - Q-\Delta)^{-1}P = (I-H\Delta)^{-1} G 
\]
where $H = (I-Q)^{-1}$ and $G$ is the original transfer function.
We seek $\Delta(s)\in \mathcal{H}_\infty$ of minimal norm so that 
$(I-H\Delta)^{-1}G$ is unbounded, thereby
destabilizing the system. We restrict ourselves to rational $\Delta$ to ensure it is a causal, time-invariant, bounded-input-bounded-output 
operator, though a thorough discussion of these topics once again eludes the current scope. 

We presently restrict ourselves to ``single-link'' perturbations, meaning that we only consider perturbations 
on the effect of one state $y_i$ on one other state $y_j$. Accordingly, the perturbation matrix $\Delta$ will have a single non-zero entry in index $(j,i)$. 
It follows from the small gain theorem (see Chapter 8 of \cite{dullerud2013course}) that the minimal norm of such a perturbation $\Delta$
that renders the system \textit{not asymptotically stable} (i.e. unstable or marginally stable) 
is $\|H_{ij}(s)\|_{\infty}^{-1}$, and any larger perturbation renders the system properly unstable. 
Hence, the $(i,j)$-th link's \textit{vulnerability} to exploitation is defined as the inverse of the minimal norm of a destabilizing perturbation: 
\[
V_{ij} = {\|H_{ij}(s)\|_{\infty}}. 
\]
Intuitively, this means a system is more vulnerable if a small perturbation can destabilize it. 
The network link of the DSF that corresponds to the largest value of $V_{ij}$ is thus the most vulnerable, as it admits the destabilizing perturbation of minimal norm.
%

%% file: 30-results.tex
\section{Numerical Results}\label{sec:results}
We propose a model of the sentiment evolution of five agents, denoted $\mathbf{x} = [x_{1} \ x_{2} \ x_{3} \ x_{4} \ x_{5}]^{T}$, subject to the influence of both one another and two distinct static sources: 
\[
\dot{x} = 
    \underbrace{\begin{bmatrix}
        -.7 & .2 & 0 & .4 & 0 \\
        .2 & -1.6 & .2 & 0 & .6 \\
        .1 & .1 & -.3 & 0 & 0 \\
        .6 & 0 & 0 & -1.6 & .4 \\
        0 & .4 & 0 & .2 & -.7 \\
    \end{bmatrix}}_{A}
    x +
    \underbrace{\begin{bmatrix}
    -.1 \\
    .4 \\
    -.1 \\
    .4 \\
    -.1
    \end{bmatrix}}_{b}.
\]
Entries $a_{ij}$ signify the influence of agent $x_{j}$ on agent $x_{i}$, and the inhomogeneous term $b$ contributes the influence of static sources.
The nonzero entries in row $i$ of $A$ signify the nodes that influence agent $x_{i}$; the nonzero entries of $A$ in column $j$ signify the nodes that are influenced by agent $x_{j}$. We treat all states as exposed ($y = 
x$) 
and further simplify by assuming that all exposed states can be both observed and manipulated. In general, these sets need not coincide.
Since our sources are represented via the time-invariant inhomogeneous term, $b$, we may denote $Bu(t) = b$.
\subsection{Perturbing an Existing Link} \label{sec:existinglink}
To begin, we consider only a perturbation to an existing link in the social network. 
Notably, this precludes the manipulation of self-links, which corresponds to an agent changing their position via an influence external to the current system (although we will proceed to consider this possibility in the following section). 
To find the most vulnerable link, we compute the vulnerability ($V_{ij} = \|H_{ij}\|_\infty$) for all existing links and conclude that the most vulnerable link has index $(5,2)$, satisfying $V_{5,2} = 1128/1051$.
This means that perturbing the influence that agent $x_{5}$ has over agent $x_{2}$ is the smallest perturbation to a single existing link that will destabilize the system. 
Accordingly, we propose
\[
\Delta(s) = \displaystyle \left[\begin{matrix}0 & 0 & 0 & 0 & 0\\0 & 0 & 0 & 0 & \frac{1051\left(1 - s\right)^{2}}{1128 \left(s + 1\right)^{2}}\\0 & 0 & 0 & 0 & 0\\0 & 0 & 0 & 0 & 0\\0 & 0 & 0 & 0 & 0\end{matrix}\right]
\]
as an appropriate additive perturbation to $Q$. 
It is minimal in $\|\cdot\|_{\infty}$ (as $1051/1128 = \|H_{5,2}^{-1}\|_\infty$), and also satisfies the additional criteria outlined in Section (\ref{sec:dsf}). 

To interpret the effect of this perturbation, we unwind the now-perturbed DSF, working backwards from Equation (\ref{eqn:dsf}) with $Q$ replaced by $Q + \Delta$. 
Since $x = y$ and $B = I$, we note that $A = A_{11} = \tilde{Q}$, $I = B = B_1 = \tilde{P}$, and $D = \text{diag}(A)$, in Equations (\ref{eqn:prepredsf})-(\ref{eqn:dsf}). 
Consequently, 
\begin{align*}
&&Y &= (Q + \Delta)Y + PU \\
&\implies &sY &= AY + (sI - D)(\Delta)Y + U \\
&&&= AY + 
\begin{bmatrix}
0 \\
\frac{1051(s+1.6)(s-1)^{2}}{1128(s+1)^{2}}Y_{5} \\
0 \\
0 \\
0
\end{bmatrix}
+ U \\
&\implies &\dot{y} &= Ay + 
\begin{bmatrix}
0 \\
\mathcal{L}^{-1} \left( \frac{1051(s+1.6)(s-1)^{2}}{1128(s+1)^{2}}Y_{5} \right) \\
0 \\
0 \\
0
\end{bmatrix}
+ u.
\end{align*}
A partial fraction decomposition yields
\begin{align*}
   d(s) &= \frac{1051(s+1.6)(s-1)^{2}}{1128(s+1)^{2}} \\
   & = .932s - 2.236 + \frac{1.491}{(s+1)} + \frac{2.236}{(s+1)^{2}}.
\end{align*}


It bears repeating that we now magnify the perturbation by a factor of $(1+\varepsilon)$ to guarantee the perturbed system has been genuinely destabilized. This is discussed further below. Taking an inverse Laplace transform of $(1+\varepsilon)d(s)$, we may now reformulate the perturbed system in the time domain. Rounding to the nearest thousandth,
\begin{align*}
    \dot{x}_{1} &= -.7x_{1} + .2x_{2} +.4x_{4} - .1 \\
    \dot{x}_{2} &= .2x_{1} - 1.227x_{2} + .2x_{3} + .187x_{4} \\
    &\phantom{=\ } - 2.291x_{5} + 1.492p + 2.238q + .4 \\ 
    \dot{x}_{3} &= .1x_{1} + .1x_{2} -.3x_{3} - .1 \\
    \dot{x}_{4} &= .6x_{1} - 1.6x_{4} + .4x_{5} + .4 \\
    \dot{x}_{5} &= .4x_{2} + .2x_{4} - .7x_{5} - .1 \\
    \dot{p} &= x_{5} - p \\
    \dot{q} &= p - q\\
\text{where }\quad
    p(t) &= e^{-t} \ast x_{5} = \int_{0}^{\infty} e^{-\tau}x_{5}(t - \tau) d\tau, \\
    q(t) &= te^{-t} \ast x_{5} = \int_{0}^{\infty} \tau e^{-\tau}x_{5}(t - \tau) d\tau.
\end{align*}
The final two expressions are convolution variables introduced for the sake of reducing the perturbed system once again to the first order. 
To verify the new system is truly unstable, we examine the perturbed matrix, $\tilde{A} \coloneq$
\[
\begin{bmatrix}
    -.7 & .2 & 0 & .4 & 0 & 0 & 0 \\
    .2 & -1.227 & .2 & .187 & -2.291 & 1.492 & 2.238 \\
    .1 & .1 & -.3 & 0 & 0 & 0 & 0 \\
    .6 & 0 & 0 & -1.6 & .4 & 0 & 0 \\
    0 & .4 & 0 & .2 & -.7 & 0 & 0 \\
    0 & 0 & 0 & 0 & 1 & -1 & 0 \\
    0 & 0 & 0 & 0 & 0 & 1 & -1
\end{bmatrix},
\]
which has spectrum $\sigma(\tilde{A}) = \{\lambda_{\varepsilon}, \allowbreak -.308, \allowbreak -.538, \allowbreak -1.625, \allowbreak -1.124 \pm 1.175 i, \allowbreak -1.809\}$. 
The presence of an eigenvalue with strictly positive real part, $Re(\lambda_{\varepsilon}) \gtrapprox 0$, guarantees an unstable system. 

By construction, the perturbation $\Delta$ is minimal so that the perturbed system is \textit{not asymptotically stable}---i.e., it moves one eigenvalue to the imaginary axis. 
However, the presence of an eigenvalue with null real part does not guarantee instability, as the system may remain marginally stable. 
To account for this, we instead implement $\Delta_\varepsilon = (1+\varepsilon)\Delta$ in our calculations and simulations to ensure the existence of a small positive eigenvalue, and hence proper instability, while still maintaining a near-minimal value in $\mathcal{H}_\infty$ norm. We choose to set $\varepsilon = .001$, though $\varepsilon$ may be taken to be as close to zero as desired. 
The simulation plotted in 
Figure (\ref{fig:comparison}) shows that, in contrast to the original model, the sentiment of all agents grows without bound. That is, targeted changes to the influence of one agent on another successfully radicalizes all agents. 


\subsection{Perturbing a Created Link} \label{sec:createdlink}
Perturbations that rely on the presence of existing links are constrained in a fundamental way: the elements of the matrix $H$ that are considered to be a candidate for minimal norm are only those indices where the matrix $Q$ is nonzero. All other elements are excluded from consideration. A created-link perturbation, on the other hand, examines the norm of all elements of $H$ to determine the destabilizing perturbation of minimal norm, including those for which the corresponding element of $Q$ is possibly null. 
Since this is a superset of the previous case, its minimal norm will be at least as small as before. 

To find the most vulnerable link, we compute the vulnerability ($V_{ij} = \|H_{ij}\|_\infty$) for all possible links and conclude that the most vulnerable link has index $(2,2)$, satisfying $V_{2,2} = 1680/1051.$
This perturbation may be interpreted as 
the susceptibility of agent $x_{2}$ to influences external to the system as presented.
We propose
\[
\Delta(s) = \left[\begin{matrix}0 & 0 & 0 & 0 & 0\\0 & \frac{1051 \left(s-1\right)^{2}}{1680 \left(s + 1\right)^{2}} & 0 & 0 & 0\\0 & 0 & 0 & 0 & 0\\0 & 0 & 0 & 0 & 0\\0 & 0 & 0 & 0 & 0\end{matrix}\right]
\]
as an appropriate additive perturbation to $Q$. We note that it is minimal in $\|\cdot\|_{\infty}$ to ensure asymptotic stability is lost, as it satisfies $1051/1680 = \|H_{ij}^{-1}\|_\infty$ as well as the additional criteria discussed in Section (\ref{sec:dsf}). 

To interpret the effect of this perturbation, we follow the unwinding procedure outlined in Section (\ref{sec:existinglink}). A partial fraction decomposition yields
%
%
%
\begin{align*}
   d(s) &= \frac{1051(s+1.6)(s-1)^{2}}{1680(s+1)^{2}} \\
   &= .626s - 1.501 + \frac{1.001}{(s+1)} + \frac{1.501}{(s+1)^{2}}.
\end{align*}
Taking an inverse Laplace transform of $(1+\varepsilon)d(s)$, we may now reformulate the perturbed system in the time domain. Rounding to the nearest thousandth,
\begin{align*}
    \dot{x}_{1} &= -.7x_{1} + .2x_{2} +.4x_{4} - .1 \\
    \dot{x}_{2} &= .535x_{1} - 8.302x_{2} + .535x_{3} + 1.605x_{5} \\
    &\phantom{=\ } + 2.681p + 4.021q + .4 \\ 
    \dot{x}_{3} &= .1x_{1} + .1x_{2} -.3x_{3} - .1 \\
    \dot{x}_{4} &= .6x_{1} - 1.6x_{4} + .4x_{5} + .4 \\
    \dot{x}_{5} &= .4x_{2} + .2x_{4} - .7x_{5} - .1 \\
    \dot{p} &= x_{2} - p \\
    \dot{q} &= p - q,
\end{align*}
where $p$ and $q$ are once again convolution variables introduced for the sake of reducing to a first order system. 
To verify a truly unstable system, we examine the perturbed matrix $\tilde{A} \coloneq$
\[
\begin{bmatrix}
    -.7 & .2 & 0 & .4 & 0 & 0 & 0 \\
    .535 & -8.302 & .535 & 0 & 1.605 & 2.681 & 4.021 \\ 
    .1 & .1 & -.3 & 0 & 0 & 0 & 0 \\
    .6 & 0 & 0 & -1.6 & .4 & 0 & 0 \\
    0 & .4 & 0 & .2 & -.7 & 0 & 0 \\
    0 & 1 & 0 & 0 & 0 & -1 & 0 \\
    0 & 0 & 0 & 0 & 0 & 1 & -1
\end{bmatrix},
\]
which has spectrum $\sigma(\tilde{A}) = \{\lambda_{\varepsilon}, \allowbreak -.320, \allowbreak -.452, \allowbreak -.706, \allowbreak -1.587, \allowbreak -1.854, \allowbreak -8.683\}$. 
As above, the presence of a an eigenvalue with $Re(\lambda_{\varepsilon}) \gtrapprox 0$, guarantees an unstable system.

We again implement a perturbation $\Delta_\varepsilon = (1+\varepsilon)\Delta$ in calculations and simulations to guarantee proper instability with $\varepsilon = .001$.
A simulation of the perturbed system is found in Figure (\ref{fig:comparison}). 
Here we see the effect of the less pronounced perturbation---the sentiment of each agent slowly grows without bound, though much more gradually than in the previous case. 
It is prudent to pause here and note just how qualitatively more subtle the created-link perturbation is when compared to the existing-link perturbation. 
This conforms to expectation. 
Further removing constraints (e.g., considering perturbation of two or more links) will result in even smaller-normed perturbations that will still destabilize the system, albeit more slowly. 

\subsection{Interpretation}
Our investigations 
on this model reveal that agent $x_{2}$ plays a critical role in regulating and directing the dynamics of the underlying social network. 
This means that a suitable alteration of the particular quality of influence characterizing agent $x_{2}$ has the potential to result in a catastrophic disruption of system dynamics. 
As the agent with the most interconnected node of any in the network, agent $x_{2}$ fulfills the role of a trendsetter, playing a pivotal role in determining the ultimate fate of this hypothetical community.
While we consider a small set of individuals' sentiments in this case study, each agent could also be taken to represent an idealized community of reasonably homogeneous sentiment. In this case, system dynamics would represent the relation and evolution of entire communities rather than individuals.
From this perspective, it's clear to see the severe consequences of such a perturbation.


Perhaps what is most interesting about these scenarios is the qualitative difference between the original equilibrium and the perturbed systems. Namely, the system transitions from a stable distribution of dissenting sentiments to a condition of all states growing without bound (as opposed to, say, one node).
This exhibits that the application of influence in subtle but precise perturbations is capable of widespread and unbounded effect on long-term system outcome.


Imagine a scenario in which the administrator of a large social network is able to measure and manipulate the influence that each user has over others, e.g., by altering the algorithm determining the social media feeds of users. 
The preceding theory could then be used to determine and implement a destabilizing perturbation to a single network link. 
For instance, a selective bias could be implemented over the content a user or group of users is presented, or one could even fabricate disingenuous content in order to manipulate user sentiment (e.g. AI generated ``fake news''). Methods to create tailored propaganda that is compelling and believable has never been more accessible, which emphasizes the need to maintain an awareness of the critical vulnerabilities of complex systems.





%% file: 31-image.tex
\begin{figure}[htbp] 
    \centering
    \includegraphics[width=0.45\textwidth]{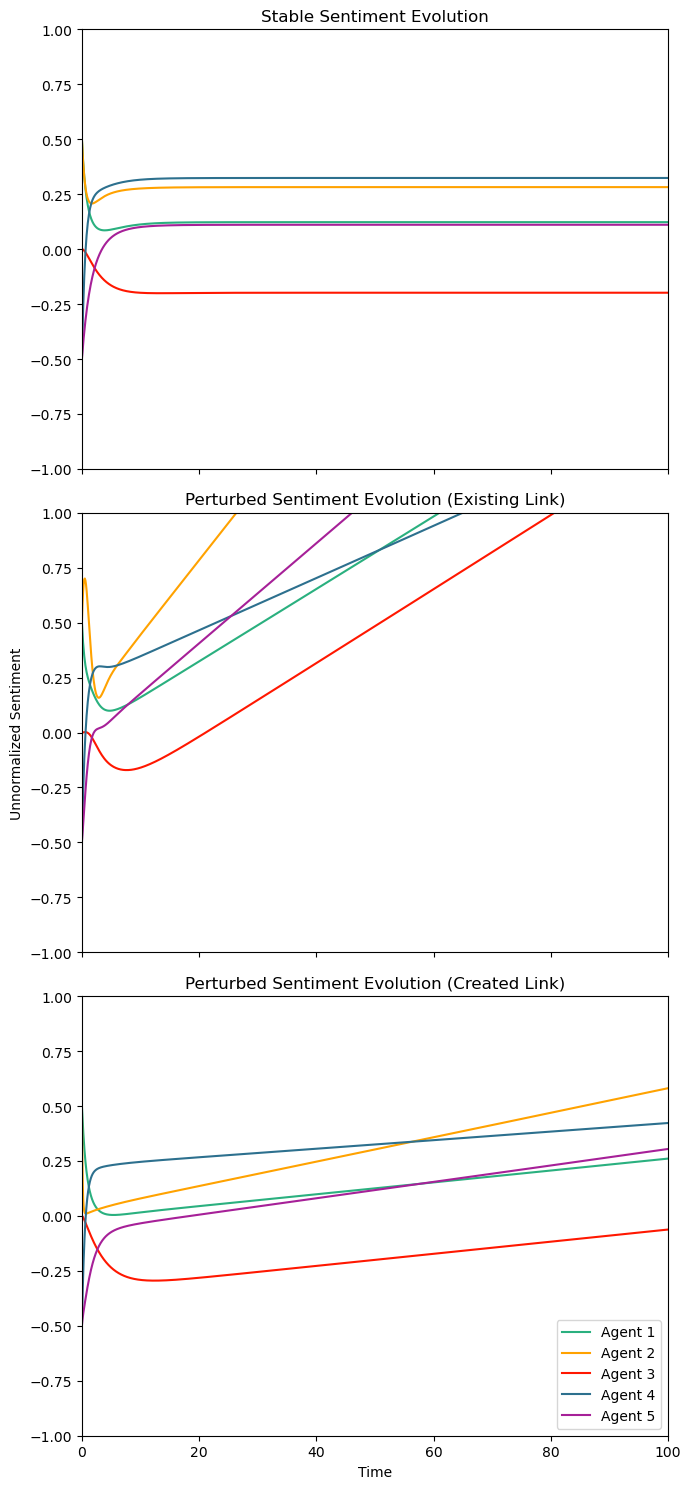}
    \caption{Plots depict a comparison of sentiment evolution of the network in Figure (\ref{fig:network}), using initial conditions for agents' sentiment $[.5 \ .5 \ 0 \ -.5 \ -.5]^{T}$: \textbf{(top)} stable system; \textbf{(middle)} system with quasi-minimally normed destabilizing perturbation to a single existent link; \textbf{(bottom)} system with quasi-minimally normed destabilizing perturbation to a single 
    existent or non-existent link. 
    In contrast to the stable evolution of the initial network (top plot), the behavior caused by the existing-link perturbation (middle plot) shows that targeted change to one agent's influence on another can radicalize all agents rapidly (see Section (\ref{sec:existinglink})); whereas, allowing perturbation of any link, existing or not, results in a more subtle effect that nevertheless also forces all agents' sentiment to grow without bound (Section (\ref{sec:createdlink})).}
    \label{fig:comparison}
\end{figure}

%% file: 40-conclusions.tex
\section{Conclusion}
In this paper, we explore the application of techniques from robust control theory to a model of social influence. To the best knowledge of the authors, this is the first time that such an analysis has been performed. In both cases considered, we found that destabilizing the network can be most efficiently achieved by applying influence to the network's most centrally connected member, agent $x_{2}$. Moreover, we discovered that in both cases of destabilization, the network dynamics trend without bound in favor of the more extreme, less broadly accepted sentiment. 
Applied to an actual social network, the propensity of these techniques for malicious use is not difficult to imagine, and underscores the necessity of responsible stewardship.

However, the model of social dynamics presented here is quite primitive. 
The implementation of a more expressive model and parameters fitted to real data would offer greater fidelity and realism.
The addition of Taylor's nonlinear model developments, the consideration of multi-link attacks, restricting the set of exposed states, and experimental verification of social network structure and parameters each seem to the authors to be fruitful topics for future research. 
Finally, we reiterate that the methods of analysis exhibited here are applicable to any similar ODE-modeled system. 
These intrinsic vulnerabilities must be evaluated and mitigated whenever possible to ensure the continued resilience of extant systems.

\section{Acknowledgments}
Thank you to Sean Warnick for his guidance throughout this project and to Ben Francis for his advice on marginal stability.
The research in this presentation was conducted with the U.S. Department of Homeland Security (DHS) Science and Technology Directorate (S\&T) under contract 70RSAT23KPM000049 and also by AI Sweden's Security Consortium and Vinnova, Sweden's Innovation Agency. 
Any opinions contained herein are those of the authors and do not necessarily reflect those of DHS S\&T. 